\documentclass[
aps,
nofootinbib,
superscriptaddress,
tightenlines,
notitlepage,
twocolumn,
showpacs,
floatfix,
]{revtex4-1}
\usepackage{amssymb,amsmath,bm,tensor,braket}
\usepackage[colorlinks]{hyperref}

\usepackage{graphicx}
\usepackage{epstopdf}
\graphicspath{{figures/}}
\usepackage{dcolumn}
\usepackage{bm}
\usepackage{xcolor}

\newcommand{\BIT}{\affiliation{School of Physics, Beijing Institute of Technology, Beijing, 100081, China}}

\begin{document}

\preprint{APS/123-QED}

\title{Neutron stars and Pulsar timing arrays as Axion giant gyroscopes}

\author{Yiming Liu}\email{7520220161@bit.edu.cn}\BIT
\author{Jinneng Luo}\email{3120221503@bit.edu.cn}\BIT
\author{Sichun Sun}\email{sichunssun@gmail.com}\BIT


\begin{abstract}
We consider the three-dimensional rotating motions of neutron stars blown by the “axion wind”. Neutron star precession and spin can change from the magnetic moment coupling to the oscillating axion background field,  in analogy to the gyroscope motions with a driving force and the laboratory Nuclear Magnetic Resonance(NMR) detections of the axion. This effect modulates the pulse arrival time of the pulsar timing arrays. It shows up as a signal on the timing residual and two-point correlation function on the recent data of Nanograv and PPTA. The current measurement of PTAs can thus cast constraints on the axion-nucleon coupling as $g_\text{ann} \sim 10^{-12}\text{GeV}^{-1}$   \end{abstract}


\maketitle


\section{Introduction}\label{sec:1}

Oscillating background fields across the cosmic space, which behave like matter, are interesting dark matter candidates \cite{Hu:2000ke,Hui:2016ltb, Arias:2012az,Feng:2010gw,RePEc:nat:nature:v:586:y:2020:i:7829:d:10.1038_d41586-020-02741-3}. The most common ones are either axionlike pseudoscalars or scalar fields such as moduli \cite{Preskill:1982cy,Abbott:1982af,Dine:1982ah,Svrcek:2006yi}. Since axion has a large parameter space, various axion detection schemes are proposed across the spectrum\footnote{https://github.com/cajohare/AxionLimits}, from the quantum laboratory and collider studies to astrophysical observations. Inspired by the nucleon spin precession detection of axion particles such as CASPEr \cite{Garcon:2017ixh}, here we extend the detection principle using the microscopic nucleon spins to the macroscopic neutron stars, such that the three-dimensional rotating motion of spinning neutron stars blown by the axion wind can yield signals for a large range of parameter space for the axion mass and decay constant.  Here in this paper we mainly discuss the case that ultralight axion wind blows on the pulsars in PTA, which yields the common spectrum on the residual with $f_c  =\frac{\omega}{2\pi}\simeq 2.42 \text{\,nHz} \left(\frac{m}{10^{-23}\text{eV}} \right)$ and two-point correlation function curve. We also briefly comment on the modulation of the pulsar period with different axion masses e.g.QCD axion $ 242 \text{\,MHz} \left(\frac{m}{10^{-6}\text{eV}} \right)$, or the resonant frequency with pulsar period $2.42 \text{\,kHz} \left(\frac{m}{10^{-11}\text{eV}} \right)$.

Recently several Pulsar Timing Arrays (PTAs) observatories including NANOGrav(North American Nanohertz Observatory for Gravitational Waves)~\cite{NANOGrav:2023gor}, Parkes PTA (PPTA)~\cite{Reardon:2023gzh}, European PTA (EPTA)~\cite{EPTA:2023gyr,EPTA:2023fyk} and
Chinese PTA (CPTA)~\cite{Xu:2023wog},
report strong evidence for gravitational wave (GW) background at the nano-Hertz waveband,
which is consistent with previous claims~\cite{NANOGrav:2020bcs, Chen:2021rqp, Goncharov:2021oub, Antoniadis:2022pcn}.
Extensive studies on potential sources of the observed 
GW background~\cite{Cang:2023ysz,Franciolini:2023pbf, 
Franciolini:2023wjm, Liu:2023ymk, Ellis:2023dgf, Wu:2023hsa, Li:2023yaj,Sun:2021yra,Li:2023vuu, Battista:2021rlh,
DeFalco:2023djo,Konoplya:2023fmh,Ben-Dayan:2023lwd,Balaji:2023ehk,Kohri:2020qqd,Inomata:2023zup,Vagnozzi:2020gtf,Benetti:2021uea,
Vagnozzi:2023lwo,Guo:2023hyp,Oikonomou:2023bli,Oikonomou:2023qfz,Choudhury:2023kam,Choudhury:2023hfm,Bhattacharya:2023ysp,Madge:2023cak,Sun:2020gem},
including supermassive black holes~\cite{NANOGrav:2023hvm,Middleton:2020asl,NANOGrav:2023hfp,EPTA:2023xxk}, 
merging PBHs~\cite{Depta:2023qst, Gouttenoire:2023nzr},
phase transitions~\cite{Bian:2020urb,NANOGrav:2021flc,Xue:2021gyq,Wang:2022wwj} and 
axion topological defects~\cite{Wang:2022rjz,Ferreira:2022zzo,Inomata:2023drn}, etc are conducted.
Nongravitational wave signals, such as the ultralight dark matter can also be observed through the timing residuals \cite{Khmelnitsky:2013lxt,Porayko:2014rfa,ULDMvPPTA:Porayko:2018sfa,Sun:2021yra,Wu:2023dnp,Omiya:2023bio,Nomura:2019cvc}. 

Most of the pulsars we observed are at the distances of order kpc from the Earth.
The de Broglie wavelength of the ultralight dark matter is around $
\lambda_{\text {dB}}=\frac{2\pi}{{m} v}
\simeq 4 \text{kpc}
\left(\frac{10^{-23}\text{eV}}{m}\right)  \left(\frac{10^{-3}}{v}\right)$. 
Fuzzy dark matter oscillates with the frequency 
$ 2.42 \text{\,nHz} \left(\frac{m}{10^{-23}\text{eV}} \right)$
 and the oscillation period $ T_c={f}_c^{-1} \simeq 13.1  \text{yr} \left(\frac{10^{-23}\text{eV}}{m}\right)$. 
 The field behaves like pressureless cold dark matter on the cosmic scale. Notably, the oscillation frequencies of fuzzy dark matter models fell into the sensitive region of PTA observations.  

\section{Axion background coupling to the pulsar magnetic moment}
	Taking into account the huge occupation number of the dark matter particle in the universe the axion field in the Galaxy can be viewed as a classical field and be approximated by\cite{Duan_2023,Khmelnitsky:2013lxt}
	\begin{gather}
		a(t, \vec{x}) \approx a_0 \cos \left(-\omega t+m_a \vec{v} \cdot \vec{x}+\phi_0\right)\label{AxionField} \\
		\omega=m_a\left(1+\frac{1}{2} v^2\right) \\
		a_0=\frac{\sqrt{2 \rho_{D M}}}{m_a}
	\end{gather}
	where $m_a$ is the mass of the axion, $v$ is the typical velocity of the axion and $\rho_{DM}\sim0.3 \mathrm{GeV/cm^3}$ is the local dark matter density.
	The Hamiltonian of the interaction between the axion and the nucleon is given by\cite{Graham_2013,Berlin:2023ubt}
	\begin{equation}
		H(t, \vec{x})=g_{a N N} \nabla a(t, \vec{x}) \cdot \vec{\sigma}\label{Hamiltonian1}
	\end{equation}
	where $g_{aNN}$ is the coulping constant and $\vec{\sigma}$ is the nuclear-spin operator.
	
	As it is known that the magnetic moment $\vec{\mu}$ can be expressed in terms of the nuclear spin $\vec{\sigma}$
	\begin{equation}
		\vec{\mu}=\gamma \vec{\sigma}\label{Gyro1}
	\end{equation}
	where $\gamma$ is the gyromagnetic ratio of the nuclear spin, related to the property of the nucleon.
Define the effective magnetic field induced by the axion\cite{Garcon_2017}:
	\begin{equation}
		\vec{B}_{A L P}(t)=g_{a N N} \sqrt{2 \rho_{D M}} \frac{\vec{v}}{\gamma} \sin \left(-\omega t+\phi_0\right)\label{pseudo-magnetic field}
	\end{equation}
	therefore we can write Eq.(\ref{Hamiltonian1}) into the form below, similar to the Hamiltonian of the particle possessing the magnetic moment in an external magnetic field: 
	\begin{equation}
		H(t)=-\vec{\mu} \cdot \vec{B}_{A L P}(t)\label{Hamiltonian3}
	\end{equation}
Now we generalize the above microscopic interaction to the macroscopic astrophysical object with a huge magnetic moment and the effective axion dark matter magnetic field. 
	
	Classically, the pulsar’s magnetic moment $\vec{\mu}$ and spin angular momentum $\vec{L}$ are given by:
	\begin{gather}
		\vec{\mu}=\frac{R^3}{2} \vec{B}_p\label{ManeticMoment} \\
		\vec{L}=I \vec{\Omega}_0=\frac{2}{5} M R^2 \vec{\Omega}_0 \label{AngularMomentum}
	\end{gather}
	where $\vec{B}_p$ is the magnetic field of the pulsar's pole, $M$ represents the pulsar's mass, $R$ stands for the radius of the pulsar and $\vec{\Omega}_0$ is pulsar's angular velocity.
	
	Substitute Eq.(\ref{ManeticMoment}), (\ref{AngularMomentum}) into Eq.(\ref{Gyro1}), and we can get the gyromagnetic ratio of the pulsar:
	\begin{equation}
		\gamma=\frac{{R^3 B_{p}}/{2}}{{2 M R^2 \Omega_0}/{5}}=\frac{5 R B_{p}}{4 M \Omega_0}\label{Gyro2}
	\end{equation}
	
	Then we can arrive at the energy change of the pulsar induced by the motion of the axion
	\begin{equation}
		H\left(t\right) =-vg_{aNN} \sqrt{2\rho_{DM}}\frac{2}{5} M R^2 \Omega_0 \hat{B}_p\left( t\right) \cdot \hat{v} \sin \left(-\omega t+\phi_0\right)\label{Hamiltonian4}
	\end{equation}
	where we have defined three unit vectors in the respective orientation$\colon\hat{B}_p\left( t\right)={\vec{B}_p\left(t\right)}/{B_p},\hat{\Omega}_0={\vec{\Omega}_0}/{\Omega_0},\hat{v}={\vec{v}}/{v}$.
	Because the magnetic axis $\hat{B}_p\left( t\right)$ rotates around the spin axis $\hat{\Omega}_0$ with angular velocity $\Omega_0$, we can decompose $\hat{B}_p\left( t\right)$ as$\colon$
	\begin{equation}
		\begin{aligned}
		\hat{B}_p\left( t\right)&=\left[\hat{B}_p\left( 0\right) -\left(\hat{B}_p\left( 0\right) \cdot \hat{\Omega}_0 \right) \hat{\Omega}_0\right] \cos{\Omega_0 t}+\\& \left[\hat{\Omega}_0 \times \hat{B}_p\left( 0\right)\right] \sin{\Omega_0 t}+\left(\hat{B}_p\left( 0\right) \cdot \hat{\Omega}_0 \right)\hat{\Omega}_0
	\end{aligned}
	\end{equation}
	Define:
	\begin{gather}
		P = \left[\hat{B}_p\left( 0\right) \cdot \hat{v}-\left(\hat{B}_p\left( 0\right) \cdot \hat{\Omega}_0 \right) \left( \hat{\Omega}_0 \cdot \hat{v}\right) \right]\label{P1} \\
		Q = \left[\left( \hat{\Omega}_0 \times \hat{B}_p\left( 0\right)\right) \cdot \hat{v}\right]\label{Q1} \\
		S = \left(\hat{B}_p\left( 0\right) \cdot \hat{\Omega}_0 \right)\left( \hat{\Omega}_0 \cdot \hat{v}\right)\label{S1}
	\end{gather}
	We then arrive at:
	\begin{equation}
		\hat{B}_p\left( t\right) \cdot \hat{v}=P\cos{\Omega_0 t}+Q\sin{\Omega_0 t}+S\label{Bdotv2}
	\end{equation}
	Assume that all the energy is converted into the rotational energy of the pulsar$\colon$
	\begin{equation}
		\frac{1}{2} I \Omega^2\left(t\right) -\frac{1}{2} I \Omega_0^2=H\left(t\right) 
	\end{equation}
 A slight change in the rotation period can be defined,
	\begin{equation}
		\Omega\left(t\right)=\Omega_0 \sqrt{1-\epsilon\left(t\right)}
	\end{equation}
Dimensionless quantity $\epsilon\left(t\right)$ then becomes 
	\begin{equation}
		\begin{aligned}
		\epsilon&\left(t\right)=\frac{2}{\Omega_0} v g_{aNN}  \sqrt{2 \rho_{D M}}\times\\&\left(P\cos{\Omega_0 t}+Q\sin{\Omega_0 t}+S\right)  \sin \left(-\omega t+\phi_0\right)
		\end{aligned}
	\end{equation}

	Next, we turn to the determination of the pulsar's timing residuals. Assume that $dt$ is the time pulsar takes to turn $d\theta$ under the action of the axion field, $dT$ is the time pulsar takes to turn the same angle without the action of the axion field, we then can determine the timing residual of the pulsar per unit time\cite{Maggiore:2018sht}$\colon$
	\begin{equation}
		z\left(t\right) =\frac{dt-dT}{dT}=\frac{dt}{dT}-1
	\end{equation}
	According to the definition of the angular velocity, we can write $dT$, $dt$ as$\colon$
	\begin{gather}
		dt=\frac{d\theta}{\Omega\left(t\right)}=\frac{d\theta}{\Omega_0 \sqrt{1-\epsilon\left(t\right)}}=\frac{dT}{\sqrt{1-\epsilon\left(t\right)}}
	\end{gather}
	$dt/dT$ can be written as:
	\begin{equation}
		\frac{dt}{dT}=\frac{1}{\sqrt{1-\epsilon\left(t \right)}}\approx1+\frac{1}{2}\epsilon\left(t\right)
	\end{equation}
	The timing residual $R(t)$ of the pulsar is defined to be the integration of $z\left(t\right)$, measured with respect to a reference time $t=0$:
	\begin{equation}
		\begin{aligned}
			R(t)&=\int_0^t d t  z(t) \\
			& =\frac{1}{\Omega_0}v g_{a N N} \sqrt{2 \rho_{D M}}\left[P \left(\frac{1}{2 \left(\omega-\Omega_0\right)} \cos \left(-\left(\omega-\Omega_0\right) t+\phi_0\right)\right.\right. \\
			& -\frac{\omega}{\left(\omega-\Omega_0\right)\left(\omega+\Omega_0\right)} \cos{\phi_0}\\&
			\left.+\frac{1}{2 \left(\omega+\Omega_0\right)} \cos \left(-\left(\omega+\Omega_0\right) t+\phi_0\right)\right) \\
			& +Q\left(\frac{1}{2 \left(\omega-\Omega_0\right)} \sin \left(-\left(\omega-\Omega_0\right) t+\phi_0\right)\right.\\&
			-\frac{\Omega_0}{\left(\omega-\Omega_0\right)\left(\omega+\Omega_0\right)} \sin {\phi_0} \\
		& \left.-\frac{1}{2 \left(\omega+\Omega_0\right)} \sin \left(-\left(\omega+\Omega_0\right) t+\phi_0\right)\right)\\&\left.+\frac{S}{ \omega}\left(\cos \left(-\omega t+\phi_0\right)-\cos {\phi_0}\right)\right]\label{R1}
		\end{aligned}
	\end{equation}
Notice that when we do the integral $\int_0^t d t  z(t)$, we do not include the case when $\omega=\Omega_0$ which has to be computed separately, and the result is$\colon$
\begin{equation}
	\begin{aligned}
		R&(t)=\frac{v g_{aNN} \sqrt{\rho_{DM}}}{\sqrt{2}\Omega_0}(P \sin{\phi_0}-Q \cos{\phi_0})t\\&+P\frac{v g_{aNN} \sqrt{\rho_{DM} }}{2 \sqrt{2} \Omega_0^2}[\cos(\phi_0-2 \Omega_0 t)-\cos{\phi_0}]\\&+Q\frac{v g_{aNN} \sqrt{\rho_{DM}}}{\sqrt{2}\Omega_0^2}\sin{\Omega_0 t}\cos(\phi_0-\Omega_0 t)\\&+S\frac{v g_{aNN} \sqrt{2\rho_{DM} }}{\Omega_0^2}[\cos(\phi_0-\Omega_0 t)-\cos{\phi_0}]\label{Reasonance}
	\end{aligned}
\end{equation}
The first term in (\ref{Reasonance}) has a linear dependence in $t$ and the timing residual will tend to infinity as time goes on. The physical meaning of this divergence is that the axion and the pulsar happen to have a resonance around $2.42 \text{\,kHz} \left(\frac{m_a}{10^{-11}\text{eV}} \right)$. and the angular velocity of the pulsar is changed permanently which can be searched accordingly. Nonetheless, given the wide spectrum of the axion and different types of pulsars, it seems plausible for the neutron star to resonate with the dark matter. 
	
	If the dark matter is the QCD axion $ 242 \text{\,MHz} \left(\frac{m_a}{10^{-6}\text{eV}} \right)$ with pulsars $\Omega_{0} \sim \frac{2\pi}{10^{-3}\mathrm{s}} \sim 6.28\times10^{3} \mathrm{Hz}$. Then we have $\omega\gg\Omega_{0} $. Then, the axion background will induce a nutation of the magnetic axis of pulsars, which might be observed by closely looking at the neutron star's movement. We will leave this and the resonant case to a future study.  \\

	For the other case, to determine the value of $R(t)$, we choose:
	\begin{gather}
		\hat{\Omega}_0=\left(0,0,1 \right)\label{Omega0} \\
		\hat{B}_p\left( 0\right)=\left(\sin{\theta},0,\cos{\theta}\right)\label{hatBp0}\\
		\hat{v}=\left(\sin{\varphi},0,\cos{\varphi} \right) \label{hatv}
	\end{gather}
	as the initial direction of the three axes.\\
	Then we have
	\begin{gather}
		P=\sin{\theta}\sin{\varphi}\label{P2}\\
		Q=0\label{Q2}\\
		S=\cos{\theta}\cos{\varphi}\label{S2}
	\end{gather}
	
	Consider a special case when $\hat{B}_p\left( 0\right)$ is parallel to $\hat{v}$, and is vertical to $\hat{\Omega}_0\colon$
	\begin{equation}
		\theta=\varphi=90^\circ
	\end{equation}

\begin{figure}[h]
\includegraphics[width=0.53\textwidth]{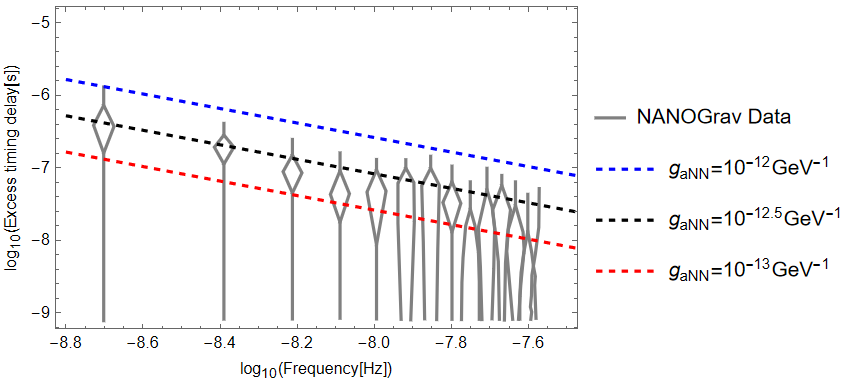}
\caption{Upper limits on the timing residual of pulsar generated by the oscillating dark matter, as a function of frequency/axion's mass. The gray line is the recently released NanoGrav data\cite{NANOGrav:2023gor}. The blue, black, and red lines are the upper limits for the coupling constant $g_{aNN}$ equals $10^{-12}\mathrm{GeV^{-1}},  10^{-12.5}\mathrm{GeV^{-1}, 10^{-13}\mathrm{GeV^{-1}}}$ respectively.}
\label{pta}
\end{figure}

For the ultralight dark matter $ 2.42 \text{\,nHz} \left(\frac{m}{10^{-23}\text{eV}} \right)$, 
then we have
	\begin{equation}
		\begin{aligned}
		max& \left| R(t,\vec{x})\right|\\&\approx 8.2\times 10^{-17}\mathrm{sec}\left(\frac{g_{aNN}}{10^{-9}\mathrm{GeV^{-1}}}\right) \left( \sqrt{\frac{\rho_{DM}}{0.3 \mathrm{GeV/cm^3}}}\right)
		\end{aligned}
	\end{equation}
Consider another special case without losing generality:
	\begin{equation}
		\theta=30^\circ \qquad \varphi=60^\circ
	\end{equation}
	The maximum value of the timing residual of the ultralight dark matter is$\colon$
	\begin{equation}
		\begin{aligned}
		max& \left| R(t,\vec{x})\right|\\&\approx 1.9\times 10^{-7}\mathrm{sec}\left(\frac{g_{aNN}}{10^{-12}\mathrm{GeV^{-1}}}\right) \left( \sqrt{\frac{\rho_{DM}}{0.3 \mathrm{GeV/cm^3}}}\right)
		\end{aligned}
	\end{equation}
This is within the reach of current PTA observation. In Fig.\ref{pta} we plot the timing residual induced by the axion together with the current Nanograv15 data. We can see that PTA data can already put constraints on the axion-nucleon coupling with some assumptions.

\section{Timing residual the two-point correlation}
\begin{figure}[h]
\includegraphics[width=0.53\textwidth]{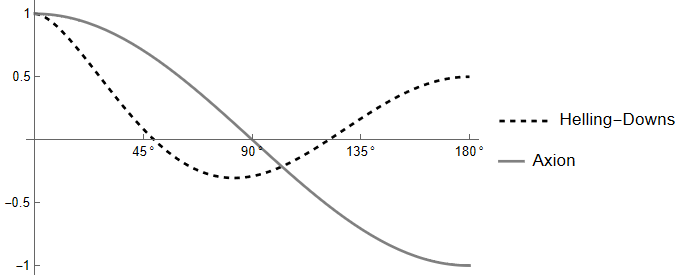}
\caption{The angle of separation dependence of the two-point correlation function, normalized to $1$ when $\theta=0$. The dashed curve corresponds to the angular correlation of the timing residuals induced by the gravitational wave background(Hellings-Downs curve) and the solid line represents the correlation of the axion's effect acting on the pulsars.}
\label{HDcurve}
\end{figure}

	The cross-correlation Hellings-Downs curve between the noise-independent data is essential to detect the gravitational wave background\cite{Hellings:1983fr,Armaleo:2020efr}. 
	The cross-correlation between the timing residuals of the pulsar a and b can be calculated through $\left\langle z_a\left(t\right)z_b\left(t\right)\right\rangle$, where $\left\langle \cdots \right\rangle$ represents a time average or an ensemble average. Having the explicit form of the redshift $z\left(t\right)$ in hand, we can then calculate,
\begin{equation}
	\left\langle z_a\left(t\right)z_b\left(t\right)\right\rangle=\frac{1}{\Omega_{0a}\Omega_{0b}}v^2 g^2_{aNN}\rho_{DM}\left\langle S_a S_b \right\rangle\cos{\left( \phi_{0a}-\phi_{0b}\right) }
\end{equation}
To simplify the calculation below, we then choose another frame that contains the pulsar location unit vector which points from the Earth towards the pulsar as the initial direction: 
\begin{gather}
	\hat{n}_a=\left(0,0,1\right)\\
	\hat{n}_b=\left(\sin{\zeta},0,\cos{\zeta}\right) 
\end{gather}
Without losing generality we set the initial direction of the magnetic axis of the pulsars a and b as:
\begin{gather}
	\hat{B}_{pa}\left(0\right)=\left(0,0,-1\right)\\
	\hat{B}_{pb}\left(0\right)=\left(-\sin{\zeta},0,-\cos{\zeta}\right)
\end{gather}
We also set the orientation of the axion's velocity and the rotational axes of pulsars a and b:
\begin{gather}
	\hat{v}=\left(\sin{\alpha}\cos{\beta},\sin{\alpha}\sin{\beta},\cos{\alpha}\right)\\
	\hat{\Omega}_{0a}=\left(\sin{\theta_a}\cos{\varphi_a},\sin{\theta_a}\sin{\varphi_a},\cos{\theta_a}\right)\\
	\hat{\Omega}_{0b}=\left(\sin{\theta_b}\cos{\varphi_b},\sin{\theta_b}\sin{\varphi_b},\cos{\theta_b}\right)
\end{gather}
According to Eq.(\ref{S1}), we can easily calculate $S$ of pulsar a and b:
\begin{equation}
	\begin{aligned}
	S_a=-\cos{\theta_a}&(\cos{\alpha}\cos{\theta_a}+\sin{\alpha}\cos{\beta}\sin{\theta_a}\cos{\varphi_a}\\&+\sin{\alpha}\sin{\beta}\sin{\theta_a}\sin{\varphi_a})
	\end{aligned}
\end{equation}
\begin{equation}
	\begin{aligned}
		S_b&=(-\cos{\zeta}\cos{\theta_b}-\sin{\zeta}\sin{\theta_b}\cos{\varphi_b})\\&\times(\cos{\alpha}\cos{\theta_b}+\sin{\alpha}\cos{\beta}\sin{\theta_b}\cos{\varphi_b}\\&+\sin{\alpha}\sin{\beta}\sin{\theta_b}\sin{\varphi_b}) 
	\end{aligned}
\end{equation}
The pulsars we observe are distributed all over the sky, we can assume that their relative angle to the direction of the axion's velocity is random. This means that the average of the cross-correlation signals among many pulsars effectively takes an average over the direction of the axion's velocity.  The rotational axis can also change its direction as a result of the motion of the galaxy, so we also take an average over the rotational axis and perform the integral over $\hat{v}, \hat{\Omega}_{0a}, \hat{\Omega}_{0b}$:
\begin{equation}
	\begin{aligned}
		\left\langle S_a S_b\right\rangle=\int\frac{d\hat{v}}{4\pi}\frac{d\hat{\Omega}_{0a}}{4\pi}\frac{d\hat{\Omega}_{0b}}{4\pi}S_a S_b=\frac{1}{27}\cos{\zeta}
	\end{aligned}
\end{equation}
This curve is similar to the vector type modes in massive gravity types of studies \cite{2008ApJ...685.1304L, Chamberlin:2011ev,Cornish:2017oic,Hellings:1983fr,Omiya:2023bio} as expected, due to the one-dimensional directional nature of this modulation. We plot this axion-nucleon curve in Fig.\ref{HDcurve}. 

\section{discussion}

We consider a new detection possibility to employ the neutron stars as the axion wind detectors. Especially current PTA network observations can provide interesting sensitivity on the axion-nucleon couplings. This classical macroscopic detection mechanism is complementary to axion microscopic production and capture mechanism \cite{Wu:2023ypz,Beznogov:2018fda,Dev:2023hax,Berlin:2023ubt,Chao:2023liu,Gao:2023gph, Caputo:2019xum} to set the limits on the related coupling. Due to the complexity of the neutron star structure models, we expect noises from several different sources to get into the neutron star motion. Nevertheless, a close analysis of neutron star behavior might provide further insight into this type of coupling.

\section*{Acknowledgements}
We thank Fei Gao for the discussion on the neutron stars. This work is supported by the National Key R$\&$D Program of China (grant 2023YFE0117200) and the National Natural Science Foundation of China (Nos. 12105013). \\

\bibliography{refs.bib}
 
\end{document}